\documentclass[12pt]{article}

\usepackage{amscd, amsmath}
\usepackage{amsthm, amssymb}
\usepackage{array}

\newcommand{\CC}{\mathbb{C}}

\newcommand{\is}{I}

\newcommand{\mais}{\varphi}

\newtheorem{thm}{Theorem}[section]
\newtheorem{prop}[thm]{Proposition}

\newtheorem{lem}[thm]{Lemma}
\newtheorem{defin}[thm]{Definition}

\theoremstyle{remark}

\theoremstyle{remark}

\theoremstyle{remark}

\newenvironment*{prooff1.1}{\noindent {\bf Proof of theorem \ref{thm1.1}.}}{\hfill $\qed$ \vspace{.3cm}}

\begin{document}

\begin{titlepage}
\title{
\vskip -70pt
\begin{flushright}
{\normalsize \ ITFA-2009-24}\\
\end{flushright}
\vskip 45pt
{\bf Non-abelian vortices, Hecke modifications and singular monopoles}
}
\vspace{3cm}

\author{{J. M. Baptista} \thanks{ e-mail address:
    j.m.baptista@uva.nl}  \\
{\normalsize {\sl Institute for Theoretical Physics} \thanks{ address: Valckenierstraat 65, 1018 XE Amsterdam, The Netherlands}
} \\
{\normalsize {\sl University of Amsterdam}} 
}

\date{July 2009}

\maketitle

\thispagestyle{empty}
\vspace{1.5cm}
\vskip 20pt
{\centerline{{\large \bf{Abstract}}}}
\vspace{.2cm}
In this note we show that for the group $G = U(N)$  the space of Hecke modifications of a rank $N$ vector bundle over a Riemann surface $C$ coincides with the moduli space of solutions of certain non-abelian vortex equations over $C$. Through the recent work of Kapustin and Witten this then leads to an isomorphism between the moduli space of vortices and the moduli space of singular monopoles on the product of $C$ with a closed interval $I$.

\vspace{.45cm}

\end{titlepage}

\section{Introduction}

In the groundbreaking and sizeable article \cite{K-W}, Kapustin and Witten, among many other insights, gave an interpretation of the Hecke modifications of a $G$-bundle over a Riemann surface $C$ in terms of singular solutions of the Bogomolny equations over the three-dimensional product $I \times C$, with $I$ a closed interval. This interpretation allowed them to establish a natural correspondence between certain Hecke operators that appear in the geometric Langlands program and certain 't Hooft operators that appear in quantum field theory, eventually leading to a far-reaching description of the mathematical Langlands duality in terms of the S-duality in QFT \cite{K-W, Fren}.

With this grand work as background and motivation, in this short note we point out that at least in the case of the group $G =U(N)$ there exists an alternative gauge theory description of the Hecke modifications of a bundle. Just as in \cite{K-W} the Hecke modifications are related to singular monopoles on $I \times C$, here we will see that they also arise naturally from non-abelian vortices over $C$. These vortices are the classical BPS-configurations of a supersymmetric ${\mathcal N}=(2,2)$ gauged linear sigma-model over the curve $C$. In particular, through the Hecke modifications, one can establish a direct relation between the vortex and the monopole moduli spaces.

We start by recalling the concept of Hecke modification in the simplest setting of $G_\CC = GL(N, \CC)$ bundles.
\begin{defin}
A Hecke modification of a holomorphic vector bundle $E_- \rightarrow C$ at the points $\{ z_1 , \ldots , z_r \} \subset C$ is a pair $(E_+ , h)$ consisting of another holomorphic vector bundle $E_+ \rightarrow C$ and an isomorphism
\begin{equation}
h :\ E_- |_{C \setminus \{ z_1 , \ldots , z_r\}} \ \longrightarrow E_+ |_{C \setminus \{ z_1 , \ldots , z_r\}} \ .
\label{1.1}
\end{equation}
In the particular case where $E_- = \CC^N \times C$ is trivial, giving $h$ is the same as giving a holomorphic trivialization of $E_+$ over the set $C \setminus \{ z_1 , \ldots , z_r\}$.
\label{defn1.1}
\end{defin}
Two Hecke modifications $(E_+ , h)$ and $(E_+' , h')$ are considered equivalent iff there exists a holomorphic isomorphism $E_+ \simeq E_+'$ that is globally defined over $C$ and takes $h$ to $h'$. Notice that this definition of equivalence does not treat the bundles $E_+$ and $E_-$ symmetrically. Another important concept is that of type of the modification at a point.
\begin{defin}
A Hecke modification is said to be of type $(m_1 , \ldots , m_N)$ at a point $z_j \in C$ when there are local trivializations of $E_-$ and $E_+$ with respect to which the isomorphism $h$ can be written in a punctured neighbourhood of $z_j$ as 
\begin{equation}
h(z)\  =\  A(z)\ {\rm diag} \big[ (z-z_j)^{m_1}, \ldots , (z-z_j)^{m_N}  \big] \ B(z) \ ,
\label{1.2}
\end{equation} 
where the square matrices $A(z)$ and $B(z)$ are holomorphic and invertible at the point $z_j$.
\label{defn1.2}
\end{defin}
By convention one usually orders $m_1 \geq \cdots \geq m_N$. Observe also that for a local Hecke modification factorized as above different choices of the matrix $A(z)$ do not change the equivalence class of the modification, for $A(z)$ can be regarded as a local automorphism of $E_+$. This does not mean, however, that the inequivalent modifications of fixed type are parameterized by $B(z)$, essentially because the factorization (\ref{1.2}) is not unique. 

As a final remark we remind the reader that the space of all Hecke modifications of a rank $N$ vector bundle at a point $z_0 \in C$ coincides with the affine grassmannian ${\rm Gr}_N$ for the group $GL(N, \CC)$, also called the loop Grassmannian (see for example \cite[ch.$\,$8]{P-S} or \cite[sect.$\,$9.3]{K-W}). The spaces of modifications of fixed type then define the so-called Schubert cells inside ${\rm Gr}_N$.

\section{Non-abelian vortex equations and moduli spaces}

The vortex equations that we have in mind arise naturally as the classical BPS-configurations of a ${\mathcal N}=(2,2)$ supersymmetric theory in two dimensions. This theory is a gauged linear sigma-model with group $U(N)$ and $N$ chiral matter fields in the fundamental representation. Since we want to consider BPS vortices over any Riemann surface $C$, the model should be topologically A-twisted. The vortex equations then appear as the instanton equations of the topological theory. From a geometrical perspective the data that we need are a hermitian vector bundle $E \rightarrow C$ of rank $N$ and degree $d$, a $U(N)$-connection $A$ on this bundle and a section $\phi$ of the direct sum of $N$ copies of $E$. Observe that the section $\phi$ can also be regarded as an ordered collection of $N$ sections of $E$, or, locally, as a function on $C$ with values on the $N \times N$ matrices that transforms in the fundamental representation of $U(N)$.  
  The vortex equations are then
\begin{align}
&\bar{\partial}^A \phi \ = \ 0    \label{2.2} \\
&\ast F_A - i e^2\, (\phi\, \phi^\dagger - \tau \, 1) \ = \ 0 \ ,\nonumber
\end{align}
where $\ast$ is the Hodge operator on the Riemann surface and $\tau$ and $e^2$ are positive constants. These particular equations were first studied in \cite{B-D-W}, and then with \cite{A-B-E-K-Y, H-T} also appeared in the physics literature. The first of the equations is invariant under the complexified (i.e. $GL(N, \CC)$) gauge transformations, whereas the second equation is only invariant under the unitary, or real, gauge transformations. Also, as usual, the second vortex equation can be formally interpreted as the vanishing condition of the moment map of the action of the group of real gauge transformations on the space of solutions of the first vortex equation, which as a space is an infinite-dimensional K\"ahler manifold. Then the usual correspondence between complex and symplectic quotients says that the space of vortex solutions modulo the real gauge transformations is isomorphic to the space of stable solutions of just the first equation modulo the complex transformations. But a solution of the first equation can be interpreted as a holomorphic structure on $E$ together with $N$ holomorphic sections, and, crucially, stability in this case means that these $N$ sections should be linearly independent on the generic fibre of $E$ \cite{B-D-W, B-D-GP-W}. Thus we see that each vortex solution determines a holomorphic structure on $E$ together with a holomorphic trivialization of the bundle over the curve $C$ minus a finite set of points, and, according to definition \ref{defn1.1}, this is precisely a Hecke modification of the trivial bundle over $C$.

$\ $

Having seen how a vortex solution naturally determines a Hecke modification, we will now give a more detailed description of the moduli space of vortex solutions. This space is to be later related to the space of Hecke modifications.

The moduli space of the particular kind of non-abelian vortices that we are considering here has been generally described in \cite{H-T, Eto-1, Bap-2, B-R}. The first two references are concerned with vortices on the complex plane;  \cite{Bap-2, B-R} also consider vortices over other surfaces, which is the case we need here.
In general, for a compact Riemann surface $C$ of large volume and for fixed rank and degree of $E$, the moduli space of solutions of (\ref{2.2}) is known to be a compact and smooth K\"ahler manifold. As a set, it can be described in the following way. 
\begin{defin}
A vortex internal structure $\is_N$ is a set of data consisting of an integer $k_0 \geq 0$ and a sequence $(V_1, \ldots , V_l)$ of non-zero proper subspaces of $\CC^N$ such that $V_{j+1} \cap V_{j}^{\perp} = \{ 0 \}$ for all indices $j=1, \ldots , l-1$. The order of the internal structure $\is_N$ is the non-negative integer $N k_0 + \sum_l  {\rm dim}_{\CC} V_l$.
\label{defn2.1}
\end{defin}
\begin{thm}[{\bf \cite{Bap-2}}]
Assume that ${\rm Vol }\, C > 2\pi (\deg E) / (e^2 \tau)$ and pick any finite set $\{ (z_1 , \is_N^1), \ldots , (z_r , \is_N^r) \}$ of distinct points on the surface $C$ and associated internal structures such that $\sum_{l=1}^{r} {\rm order}( \is_N^l) = {\rm degree}\ E$. Then there is a solution $(A,\phi)$ of the non-abelian vortex equations (\ref{2.2}), unique up to gauge equivalence, such that $\det \phi$ has zeros exactly at the points $z_j$ and $\phi$ factorizes around each $z_j$ with internal structure $\is_N^j$. Furthermore, all solutions of (\ref{2.2}) with $\det \phi$ not identically zero are obtained in this way.  
\label{thm2.1}
\end{thm}
The statement that the matrix $\phi$ factorizes around a point $z_0 \in C$ with internal structure $I_N = (k_0 , V_1 , \ldots, V_l)$ just means that, in a neighbourhood of $z_0$, one can write uniquely
\begin{equation}
\phi (z) \ = \ A(z) \ (z-z_0)^{k_0} \ T_{V_l}(z-z_0) \ \cdots \ T_{V_1} (z-z_0) 
\label{2.3}
\end{equation}
for some matrix $A(z)$ that is holomorphic and invertible around $z_0$. Here the matrix functions $T_V (z- z_0)$ are defined as the linear maps 
\begin{equation*}
T_V (z -z_0) \ := \ (z-z_0) \Pi_V + \Pi_V^\perp \ :\  \CC^N \longrightarrow \CC^N
\end{equation*}
constructed with the help of the natural projections associated with an orthogonal decomposition $\CC^N = V \oplus V^\perp$. In order to define these decompositions, and hence the maps $T_V$, we need of course to choose and fix a hermitian product on $\CC^N$. Theorem \ref{thm2.1} is then valid for any such choice. Observe that this product on $\CC^N$ is not in any way related to the hermitian product on the fibres of the hermitian vector bundle $E \rightarrow C$.

Looking at the definition above, it is clear that $T_V (z-z_0)$ has kernel $V$ at $z_0$ and determinant $(z-z_0)^{\dim V}$. This implies in particular that $\det \phi (z)$ vanishes at $z_0$ with multiplicity equal to the order of $I_N$.
The non-intersection condition in the definition \ref{defn2.1} of vortex internal structure appears because we want the factorization (\ref{2.3}) above to be unique. If we had allowed the vector spaces $V_j$ to be arbitrary, it would follow for instance from the identity
\begin{equation}
T_{V}(z-z_0) \ T_{V^\perp} (z-z_0) \ = \ (z-z_0)^{\dim V}
\label{2.4}
\end{equation}
that the uniqueness could not hold.

\section{Vortex  internal structures and Hecke modifications}

The aim of this section is to show that the spaces of Hecke modifications at a point and the spaces of vortex internal structures are one and the same object. This identification provides a very palpable description -- in terms of finite collections of vector spaces -- of the affine Grassmannian and its Schubert cells.  

Firstly, comparing the local factorizations (\ref{1.2}) and (\ref{2.3}), it is clear that any holomorphic matrix $\phi (z)$ described by (\ref{2.3}) determines a Hecke modification with $m_1 , \ldots, m_N \geq 0$. As we will see later, the values of the integers $m_j$ are related to the dimensions of the different   subspaces $V_j$. Moreover, since for both vortices and modifications the equivalence relation is given by the different possible choices of the invertible matrix $A(z)$, the uniqueness of (\ref{2.3}) shows that the degrees of freedom of inequivalent Hecke modifications are precisely parameterized by the choices of vector spaces $V_j$, i.e. by the vortex internal structure $I_N$. The conclusion is then that the space of all local Hecke modifications with finite and non-negative $m_j$'s is the same as the space of all different internal structures $I_N$.

$\ $

 To make the correspondence more precise, a  natural question is to ask what internal structures $I_N$ correspond to modifications of fixed type $m_1 \geq \cdots \geq m_N \geq 0$. Observe that the space of internal structures is the disjoint union of strata labeled by the integers $k_0 ,\,  {\rm dim}\, V_1 \geq  \cdots  \geq  {\rm dim}\, V_l$, and that this is analogous to the stratification of the space of Hecke modifications by the type of the modification. So it is natural to expect the integers $m_j$ to be related to the dimensions of the $V_j$'s. In fact, a careful look at the factorizations (\ref{1.2}) and (\ref{2.3}) and at the way internal structures emerge in \cite{Bap-2}, shows that the Hecke modifications of type $(m_1 , \ldots , m_N)$ correspond exactly to the internal structures with
\begin{align}
k_0\  &=\  m_N     \label{3.1}     \\
{\rm dim}\, V_k \ &= \ \# \{ j :\  m_j - m_N - k \geq 0 \}  \nonumber \ .
\end{align}  
This formula is a dictionary between the type of the modification and the type of the vortex internal structure.
Now, it is well-known that the spaces of Hecke modifications of fixed type are not in general compact. Moreover, in \cite[Sect.\,9.2]{K-W}, the authors describe how it is possible to obtain a natural compactification of these spaces by completing them with modifications of different type. Thus another reasonable question is to ask how these compactifications look like in the vortex picture, and we now turn to this.

Firstly, suppose that one fixes the type of the vortex internal structure, i.e. that one fixes the dimensions of the subspaces $V_j$ and only allows them to turn around in $\CC^N$. If all the different orientations of the $V_j$'s were allowed, then the resulting space of internal structures would certainly be compact -- it would be a product of Grassmannians. This, however, is not the case, because certain orientations are excluded by the non-intersection condition imposed in definition \ref{defn2.1}, and so the spaces of internal structures of fixed type are also in general non-compact. And what about their natural compactification? One possible answer would be just to complete these spaces with the forbidden orientations of the $V_j$'s, and hence compactify them topologically as a product of Grassmannians. If we do this, however, we are not completing the spaces with genuine internal structures of different type, because formula (\ref{2.4}) and lemma \ref{lem3.2} say that there is an equivalence relation among the forbidden orientations of the $V_j$'s. This means that this naive compactification cannot be the one described in \cite{K-W}. Instead, looking carefully at the definitions in \cite[Sect. 9.2]{K-W} and using the dictionary (\ref{3.1}) to translate the results, one can check that the natural compactification of the space of internal structures of type $(k_0 , \dim V_1 ,  \ldots , \dim V_l)$ is obtained by adding to it all the other internal structures $(k'_0 , V'_1 ,   \ldots , V'_{l'})$ of the same order that satisfy
\begin{equation}
\sum_{r\geq j} {\rm dim}\, V'_r \ \leq \ \sum_{r\geq j} {\rm dim}\, V_r  \qquad \qquad {\rm for\ all\ }\ j \in {\mathbb N}. 
\label{3.2}
\end{equation}
This is the translation to the vortex picture of the compactification of the space of Hecke modifications of type $(m_1 , \ldots , m_N)$. It actually corresponds to adding the forbidden orientations of the $V_j$'s and then quotienting by the equivalences determined by lemma \ref{lem3.2}. This quotient is encapsulated in the map $\varphi$ of the proposition below.
\begin{prop}
For $m_j \geq 0$, the space ${\mathcal Y}(m_1 , \ldots , m_N)$ of Hecke modifications of fixed type is the same as the space ${\mathcal I_N}(k_0 , \ldots , k_l)$ of vortex internal structures with dimensions $k_j = \dim\, V_j$ determined by (\ref{3.1}). Moreover, the natural compactification $\overline{{\mathcal Y}}(m_1 , \ldots , m_N)$ described in \cite[sect.\,9]{K-W} coincides with the space $\overline{{\mathcal I}}_N (k_0 , \ldots , k_l)$ described above in (\ref{3.2}). Finally, these identifications show that there exists a natural surjective map
\begin{equation}
\mais \ :\ {\rm Gr}\, (k_1 , N) \times \cdots \times {\rm Gr}\, (k_l , N) \ \longrightarrow \ \overline{{\mathcal Y}}(m_1 , \ldots , m_N)
\end{equation}
that is one-to-one on the inverse image of ${\mathcal Y}(m_1 , \ldots , m_N)$, which is an open dense subset of the domain.
\label{prop3.1}
\end{prop}
The construction of the map $\mais$ is quite simple and goes as follows. Let any collection of subspaces $L = (V_1 , \ldots, V_l)$ be a point in the domain of $\mais$.  Irrespective of whether these subspaces satisfy or not the non-intersection condition in definition \ref{defn2.1}, it makes sense to consider the the linear transformations
\begin{equation*}
\hat{T}_L (z) \ := \ T_{V_l} (z)  \cdots T_{V_1}(z)
\end{equation*} 
for all complex $z$. It is clear that $\hat{T}_L (z)$ is holomorphic and that $\det \hat{T}_L (z)$ has a single zero at $z=0$ of order $\sum_j \dim\, V_j$. In particular, applying to $\hat{T}_L (z)$ the factorization (\ref{2.3}), there is a unique internal structure $\is_N = (k_0 , V'_1 , \ldots,  V'_{l'})$ of order $\sum_j \dim V_j$ such that 
\begin{equation*}
\hat{T}_L (z) \ = \ z^{k_0}\: T_{V'_{l'}}(z) \cdots T_{V'_1}(z) \ .
\end{equation*}
So we just define $\mais (L) =  \is_N$. Observe that if the subspaces $V_j$ in the original collection $L$ do satisfy the non-intersection condition of definition $\ref{defn2.1}$, which is the generic case, then they already define a vortex internal structure, and so by the uniqueness of the factorization (\ref{2.3}) necessarily $\mais (L) =L$. If on the other hand they do not satisfy the condition, then the factorization (\ref{2.3}) will produce a different collection of subspaces $\mais (L) = (k_0 , V'_1 , \ldots,  V'_{l'})$ that defines a genuine internal structure, which can be recursively determined through lemma \ref{lem3.2} below. One can subsequently check that this new internal structure belongs to the compactification $\overline{{\mathcal Y}}(m_1 , \ldots , m_N)$, and that all internal structures in this compactification can be obtained in this way (for more details, including the proof of the lemma, see \cite[Sect. 4.1]{Bap-2}).

\begin{lem}
Let $V_1$ and $V_2$ be any two subspaces of $\CC^N$. Then calling $W$ the intersection $V_2 \cap V_1^\perp$, the linear transformations $T_V (z)$ defined in section 2 satisfy the algebraic identity
\begin{equation}
T_{V_2}(z)\ T_{V_1}(z) \ =\ T_{V_2 \, \cap \, W^\perp }(z) \  T_{W \oplus V_1} (z) \ .
\end{equation}
Observe moreover that the two subspaces on the right-hand side satisfy the usual non-intersection condition, i.e. $V_2 \cap W^\perp$ has zero intersection with $(V_1 \oplus W)^\perp$.
\label{lem3.2}
\end{lem}

A last observation, before we give two examples, is to note that as long as one is only interested in the spaces of modifications ${\mathcal Y}(m_1 , \ldots , m_N)$ and not on the modifications themselves, the restriction of having $m_j \geq 0$ is not a serious one. This is so because the spaces ${\mathcal Y}(m_1 , \ldots , m_N)$ are invariant under a simultaneous shift $m_j \mapsto m_j + a$ for any integer $a$, as explained for example in \cite{K-W}. One way to recognize  that this is true is to consider the standard line-bundle ${\mathcal O}(z_0)$ over $C$ with its standard section $s_{z_0}$ (which has a simple zero at $z_0$) and, going back to definition \ref{defn1.1}, to note that the map $(E_+ , h)  \mapsto (E_+ \otimes {\mathcal O}(z_0)^a ,  \, h \, s_{z_0}^a )$ defines an invertible correspondence between the Hecke modifications of the trivial bundle of type $(m_1 , \dots , m_N)$ and the modifications of type $(m_1 +a , \dots , m_N + a)$.

  $\ $

Two particular cases of the identifications in proposition \ref{prop3.1} are the following. 
\begin{itemize}

\item To Hecke modifications of type $(a,0, \ldots , 0)$ correspond internal structures with $k_0 = 0 $ and $a$ unidimensional vector spaces $V_1 , \ldots , V_a$ of $\CC^N$. The compactification of this space is the space of all possible internal structures $I_N$ of order $a$, and the map $\mais$ takes the product of $a$ copies of $\CC {\mathbb P}^{N-1}$ onto this compactification.

\item To Hecke modifications of type $(1,\ldots,1,0, \ldots , 0)$, with $k$ ones, correspond internal structures with vanishing $k_0$, with a $k$-dimensional $V_1$, and all other vector spaces equal to zero. These internal structures are parameterized by the Grassmannian ${\rm Gr}\, (k, N)$, which is already compact, and $\mais$ is in this case the identity map.
\end{itemize}

Expanding a bit on the first example, it is obvious that for $a=1$ the space of all vortex internal structures of order $a$ is just $\CC {\mathbb P}^{N-1}$. For $N= a=2$ the same space was shown in \cite{Eto-2} to be the weighted projective space $W \CC{\mathbb P} (1,1,2)$, which coincides with the corresponding compactified space of modifications described in \cite{K-W}. For $a=2$ and arbitrary $N$ the spaces of internal structures can be pictured as the product $\CC {\mathbb P}^{N-1} \times  \CC {\mathbb P}^{N-1}$ with the diagonal collapsed into a Grassmannian ${\rm Gr}(2, N)$ or, alternatively, as a nice finite-dimensional quotient of a space of matrices \cite{Bap-2, Eto-2}.

\section{Vortices and singular monopoles}

\subsection{Identification of moduli spaces}

Sections 9 and 10 of \cite{K-W} give an interpretation of the Hecke modifications of a $G$-bundle in terms of  a three-dimensional gauge field theory. More precisely, they describe an isomorphism between the spaces of  modifications of fixed type and a moduli space of solutions of the Bogomolny equations over the product $[t_- , t_+] \times C$ with prescribed boundary conditions and singularities (see also \cite{Nor, C-H}). In this setting, and in our case $G= U(N)$, one considers a hermitian vector bundle $E$ defined over the manifold $[t_- , t_+] \times C$ minus a finite set of interior points $p_i = (t_i , z_i)$, and takes as fields a $U(N)$-connection $A$ on $E$ and a section $\Phi$ of the adjoint bundle ${\rm ad}\, E$. The restriction of the bundle $E$ to the boundary curve $C_- = \{ t_-\} \times C$ should be trivial, whereas the restriction to $C_+$, which we call $E_+$, need not be. One then looks at the standard 3-dimensional Bogomolny equations 
\begin{equation}
F_A \ = \ \ast \nabla^A \Phi
\label{4.1}
\end{equation} 
for the pair $(A, \Phi)$ on $I \times C \setminus \{ p_i \}$ with the boundary conditions that: $(i)$ both $A \: |_{C_-}$ and $\Phi \: |_{C_+}$ should be trivial and $(ii)$ near each singularity $p_i$ the Higgs field should in some gauge be asymptotically of the form 
\begin{equation}
\Phi (p) \ \sim  \ \frac{\sqrt{-1}}{2\ {\rm distance}(p, p_i)}  \: {\rm diag}\: [m_1^i , \ldots , m_N^i ] \ .
\label{4.2}
\end{equation}
The relation between this gauge theory problem and the Hecke modifications can then be roughly encapsulated in the following result.
\begin{prop}[{\bf \cite{K-W, Nor}}]
Consider the space of solutions of the Bogomolny equations (\ref{4.1}) subject to the boundary conditions $(i)$ and $(ii)$, modulo gauge transformations on $E$ that are trivial  on $C_-$. If the singularities $p_i = (t_i , z_i)$ all have distinct $z_i$'s, then this moduli space is naturally isomorphic to the space of inequivalent Hecke modifications of the trivial bundle $\CC^N \times C$ that have type $(m_1^i , \ldots , m_N^i)$ at the respective point $z_i \in C$.
\label{prop4.1}
\end{prop}
On the other hand, when the integers $m_j^i$ are all non-negative, we have seen in the previous section that each space of Hecke modifications of fixed type coincides with a space of vortex internal structures $I_N$ with subspaces $V_j \subset \CC^N$ of fixed dimensions. This then means that the Bogomolny moduli space of the proposition above is just a cartesian product of spaces of internal structures with fixed dimensions. Furthermore, from section 2 we also know that a choice of a finite number of pairs $(z_i , I_N^i)$ is the same thing as a point in the moduli space of vortex solutions. The conclusion is thus that the vortex and monopole moduli spaces are the same.
\begin{prop}
Assume that the integers $m_j^i$ are all non-negative. Then the monopole moduli space of proposition \ref{prop4.1} is isomorphic to the vortex moduli space of solutions $(A, \phi)$ of (\ref{2.2}) on the bundle $E_+ \rightarrow C_+$ such that the matrix $\phi$ factorizes around the points $z_i \in C$ according to internal structures $I_N^i$ with dimensions fixed by the relations (\ref{3.1}).
\label{prop4.2}
\end{prop}

\begin{prop}
Consider the monopole moduli spaces of proposition \ref{prop4.1}, but leaving the type and the precise location of the singularities unspecified, so that these are additional moduli. Demand nonetheless that the singularities have distinct $z_i$'s and that their type satisfies $m_j^i \geq 0$. Then the resulting monopole moduli space is a smooth complex manifold of dimension $N\, \deg (E_+)$ that is isomorphic to the moduli space of vortex solutions on $E_+ \rightarrow C_+$ described in theorem \ref{thm2.1}.
\label{4.3}
\end{prop}
These identifications of vortex and monopole moduli spaces have been made through the picture of the moduli spaces as complex quotients of spaces of stable solutions. Firstly, the holomorphic structure on $E_+ \rightarrow C_+$ induced by a monopole connection $A$ is used as part of a unique stable pair $(A \: |_{C_+} ,\: \phi)$ that solves the first vortex equation. Secondly, the usual the Hitchin-Kobayashi correspondence says that there exists a complex gauge transformation on $E_+$ that takes this stable pair to a full vortex solution of (\ref{2.2}). The conclusion is that the monopole connection $A$ is related by a complex gauge transformation on $C_+$ to a unique vortex connection, up to real gauge transformations, and so the moduli of the monopole theory on $I \times C$ are all encapsulated in the vortex theory on the boundary $C_+$.

\section{Open questions}

\subsection{Are vortices boundaries of singular monopoles?}

The identifications of vortex and monopole moduli spaces described in the last section implied that the monopole connection $A$ is related by a complex gauge transformation on $C_+$ to the corresponding vortex connection. Given this fact, it is natural to ask if there is a more direct correspondence between the monopole solutions and the vortex solutions, i.e. if the latter complex gauge transformation can in fact be a real transformation.  To see what this means, start by breaking up the Bogomolny equation (\ref{4.1}) into components as in \cite{K-W}:
\begin{align}
(F_A)_{z \bar{z}}  \ &= \ \nabla^A_t  \Phi  \ ( \omega_C)_{z\bar{z}}   \label{5.1}  \\
(F_A)_{t z} \ &= \  i \, \nabla^A_z \Phi    \nonumber \\
(F_A)_{t \bar{z}} \ &= \ -i\, \nabla^A_{\bar{z}} \Phi  \ ,  \nonumber
\end{align}
where $\omega_C$ is the K\"ahler form on the Riemann surface.
Comparing (\ref{2.2}) and (\ref{5.1}), the question is therefore if given a vortex solution $(A, \phi)$ there is a corresponding monopole solution satisfying the additional boundary condition
\begin{equation}
\nabla^A_t \Phi \ \stackrel{?}{=} \ i e^2\, (\phi\, \phi^\dagger - \tau \, 1)       \qquad {\rm on} \ \ C_+ \  . 
\label{5.2}
\end{equation}
This monopole solution would then necessarily be unique up to real gauge transformations that are trivial on both boundaries.

 In general one should not expect (\ref{5.2}) to be true. One reason is that in the monopole moduli space of proposition \ref{prop4.1} the coordinates $t_i$ of the singularities are not specified, and different choices of $t_i$ should produce monopole solutions which are not real gauge equivalent on $C_+$, and so cannot all be real gauge equivalent to any given vortex solution. It is tempting, nevertheless, to imagine that if we are considering a single monopole singularity, or if all the $t_i$'s are chosen with the same value -- say equal to $t_0 \in (t_- , t_+)$ -- then the answer to (\ref{5.2}) could be affirmative. In this case the only value of the parameter $\tau$ for which this could possibly be true would be a function of the position of $t_0$ in the interval  $(t_-, t_+)$, and it would be interesting to find what this function is. An alternative approach\footnote{suggested by Nuno Rom\~ao.}, would be to scrap the original Dirichlet boundary condition $\Phi \: |_{C_+} = 0$ of section 4, substitute it by the inhomogeneous Neumann condition (\ref{5.2}), and then see if each vortex solution on $C_+$ determines in this way to a unique monopole solution on the bulk with the prescribed singularities.

% In any case, even if (\ref{5.2}) is not true as it stands, the fact that there exists a complex gauge transformation that relates both sides of the equation means that we can choose a unique hermitian metric on $E_+ \rightarrow C_+$ that produces vortices $(A, \phi)$ which do satisfy (\ref{5.2}). (Recall that the vortex solutions depend on the hermitian metric on $E_+$ through the hermitian conjugate section $\phi^\dagger$ that appears in the equations.) This other metric on $E_+$ would not in general coincide with the restriction to $E_+$ of the original hermitian metric on $E$. It -- meaning the metric or the complex gauge transformation, as one prefers -- would relate the vortex and the monopole solutions, and would also depend on the values of the parameters $t_i$ and $\tau$. It could prove to be an interesting object to study.

\subsection{Meromorphic vortices}

  Another natural question is whether it is possible to eliminate from proposition \ref{prop4.2} the bothersome condition $m_j^i \geq 0$. As explained in \cite{K-W} and in section 3, as long as one is only interested in the spaces of modifications and not in the modifications themselves, or in the monopole moduli spaces and not in the monopole solutions themselves, this restriction of having non-negative $m_j$'s does not lead to any loss of generality, since all these spaces are isomorphic under a simultaneous shift of the $m_j$'s by an arbitrary constant integer. However, when one starts trying to relate concrete monopole solutions and Hecke modifications to concrete vortex solutions on $C_+$, it is clear that this is not directly possible with negative $m_j$'s, because so far we have assumed that in a vortex solution the section $\phi$ is holomorphic, and comparing (\ref{1.2}) with (\ref{2.3}), a Hecke modification with negative $m_j$'s clearly demands that we allow vortex solutions with meromorphic $\phi$ as well. 
  Thus one possible strategy to eliminate the restriction on the $m_j$'s is just that: to consider meromorphic solutions of (\ref{2.2}) and then work to see if the usual correspondence between full vortex solutions and stable solutions of $\bar{\partial}^A \phi = 0$ still holds in the meromorphic setting. An alternative method would be to stick with the manifestly holomorphic sections and assume instead that $\phi$ has values in some compactification of the space of $N\times N$ matrices. One can then subsequently use the Hitchin-Kobayashi correspondence of \cite{Mun} to relate stable sections with vortex solutions. The latter method works at least in the simple $G = U(1)$ example. In this case $E_+$ is a principal $U(1)$-bundle and the target of the section $\phi$ is taken to be  $\CC {\mathbb P}^1$ with the standard circle action. Then the moduli space of solutions of the corresponding vortex equations is basically just the space of divisors on $C_+$ \cite{Yang, Bap-1}, and, as desired, it coincides with the space of Hecke modifications of the trivial line-bundle on $C$, with no restriction on the $m^i$'s. 

Understanding the vortices related to Hecke modifications with arbitrary $m_j^i$ is also relevant to the vortex/modification correspondence for groups $G$ other than $U(N)$. For instance the Hecke modifications for $G = SU(N)$ are the same as those for $U(N)$ with just the additional constraint $\sum_j m_j = 0$, which if it is to be non-trivial requires also negative $m_j$'s. Vortices with different groups $G$ have been studied for example in $\cite{Ban, Mun, Bap-1, Eto-3}$.

\subsection{Hanany-Tong construction versus Nahm equations with singularities}

The relations between the vortex and monopole moduli spaces described in section 4 ought to be visible as well after performing ADHM-like transformations. For the case of non-abelian vortices on the complex plane $C = \CC$, in the article \cite{H-T} Hanany and Tong used D-branes to produce a description of the vortex moduli spaces as finite-dimensional symplectic quotients of matrix spaces, a result analogous to the ADHM quotients for instantons. This description was then more explicitly related to the vortex solutions in \cite{Eto-1}.  On the other hand it is well-known that monopoles with singularities can be described by singular solutions of Nahm's equations \cite{C-D, Kron, Wit}. It then seems natural to ask whether one can directly relate the symplectic quotients of \cite{H-T} to the moduli spaces of singular solutions of Nahm's equations.

%\subsection{Hecke modifications of Higgs bundles}

%In the case of Hecke modifications of the trivial bundle $E_- \rightarrow C_-$, a relevant object to study was the set of holomorphic structures on $E_+$ together with $N$ holomorphic sections. Analogously, in the case where $E_-$ is not trivial, the relevant object should be the set of holomorphic triples. A holomorphic triple $(E_+,E_-, \phi)$ is composed of two holomorphic vector bundles together with a holomorphic homomorphism $\phi : E_- \rightarrow E_+$, and there is a notion of stable triple. When there is a non-vanishing Higgs field and $(E_ , \varphi)$ is a Higgs bundle, one should presumably demand that $\phi \, \varphi \, \phi^{-1}$ is also holomorphic.   

\vskip 25pt
\noindent
{\bf Acknowledgements.}
I would like to thank Nuno Rom\~ao for comments on the first version of this manuscript.  I am partially supported by the Netherlands Organisation for Scientific Research (NWO) through the Veni grant 639.031.616.

%\subsection{Langlands duality ?}

%Could the interpretation of Hecke modifications in terms of vortices and gauged linear sigma-models help in any way to understand Langlands duality ?

 %The vortex solutions are the classical instanton configurations of an A-twisted version of the ${\mathcal N}= 2$ gauged linear sigma-model. There are conjectures about non-abelian mirror symmetry (T-duality) between GLSM's and Landau-Ginzburg models. Could this have anything to do with Langlands duality ? 

\end{document}